\documentclass[12pt]{article}
\usepackage{amssymb}
\usepackage{amsmath}
\usepackage{mathrsfs}
\usepackage{commath}
\usepackage{graphicx}
\usepackage{titlesec}
\usepackage[utf8]{inputenc}
\usepackage{setspace}
\usepackage[margin=0.80in]{geometry}
\usepackage{authblk}
\usepackage{listings}
\usepackage{multirow}
\usepackage{subfig}
\usepackage{amsthm}
\theoremstyle{definition}

\begin{document}
\title{Perception and Attitude of Reddit Users Towards Use of Face-Masks in Controlling COVID-19}
\date{}
\author[1]{G.N. Singh} 
\author[2$^*$]{D. Bhattacharyya} 
\author[3]{A. Bandyopadhyay}
\affil[1]{\footnotesize Department of Mathematics \& Computing,
	Indian Institute of Technology (Indian School of Mines),
	Dhanbad-826 004, Jharkhand, India. Email: gnsingh\_ism@yahoo.com, gnsingh@iitism.ac.in }
\affil[2]{\footnotesize Department of Mathematics \& Computing,
	Indian Institute of Technology (Indian School of Mines),
	Dhanbad-826 004, Jharkhand, India. Email: diya\_bhattacharyya@yahoo.co.in 
}  
\affil[3]{\footnotesize Department of Mathematics, Asansol Engineering College, Asansol-713305, India. \newline  Email: arnabbandyopadhyay4@gmail.com}
	
\maketitle
	
\begin{abstract}
In the wake of the COVID-19 pandemic, World Health Organization (WHO) recommended the use of face-masks to combat the disease. A fraction of the population has been vocal about their disapproval, especially on social media. Motivated by the need to understand their opinions, an international online survey in English was shared to the community /r/SampleSize in Reddit. The results obtained were summarized; binomial and multinomial confidence intervals were constructed for variables of interest; Pearson's Chi-squared test was used to test whether relationship exists between two variables of interest. A second, smaller survey was conducted to cross-validate the results. The study demonstrates that despite
the digital polarization of public opinion, opinions from both sides of the spectrum are not
ubiquitous.
\end{abstract}

\textbf{Keywords}: Attitudes, COVID-19, Data analysis, Health, Opinions, Public opinion,  Sampling, Survey Research.\\

\textbf{MSc Subject Classification:} 62D05, 62-07.

\section{Introduction}
\label{Intro}

In the final months of 2019, the world witnessed the outbreak of a novel disease in the city of Wuhan in the Hubei province of China. The source was determined to be Coronavirus disease 2019 (COVID-19), caused by SARS-CoV-2 (\cite{wang}). As of 3rd April 2021, as per World Health Organization (WHO) records (\cite{who}), 129,902,402 cases of COVID-19 have been confirmed, with 2,831,815 people succumbing to death. Many countries have been gravely affected, including the United States of America, Brazil, India, Russian Federation, France, the United Kingdom, Italy, Spain, Turkey, Germany, etc. \\ 

Several health measures have been recommended by WHO to curb the spread of the disease. Such safety protocols include maintaining interpersonal distance, ensuring rooms are well ventilated, avoiding crowded places, maintaining personal hygiene by sanitizing hands regularly, using personal protective equipments like masks, etc. (\cite{whomask2}, \cite{whomask}). 
The focus of the present study is on the perception and attitude of the global population towards the utilization of masks to combat the pandemic, expressed via social networks. Since social networks vary widely in demographics as well as content. one umbrella study covering all such networks is insufficient and unlikely to properly illuminate the situation. Hence, the scope of the study has been kept narrow and limited to one specific subreddit.\\

\subsection{Motivation for current study and some relevant recent works:}

The study has been motivated by the various anti-mask protests that have erupted across the world, in countries like USA, UK, Canada, Italy, Ireland, etc. (\cite{antimask_uk},  \cite{antimask_dublin}, \cite{antimask}, \cite{antimask_us}). The public demonstrations are a stern indication that a fraction of the population oppose public health initiatives such as covering the face with protective masks. Hence, it is prudent to analyze the viewpoints of the people on the matter. \\

In the current socio-economic and political scenario, social media plays an important role in the propagation of information as well as misinformation about various relevant topics. \cite{sm1}, \cite{sm2}, \cite{sm3}, among others, have discussed the role of social media as a platform for advocacy and information, especially during crisis like mass emergencies. Many, such as \cite{echo1}, \cite{echo2}, \cite{echo3}, \cite{echo4}, have noted an ``echo chamber" effect in social media platforms, wherein similar-minded users form their own clusters or communities of individuals adhering to the same set of beliefs. Others like \cite{twittertoxicity} have argued that opinions of vocal minorities can become amplified on Twitter by use of verbal aggression such as toxic language. \\

COVID-19 is a global pandemic affecting the lives of millions across the continents. Individuals have engaged in wide-spread discussions related to COVID-19 across social media platforms such as Facebook, Twitter, Instagram, Reddit, etc. News articles in various websites, such as \cite{article1}, \cite{article2}, \cite{article3}, \cite{article4}, have highlighted the anti-mask movement that has spread across social networks in the wake of the pandemic. The rise of such anti-mask sentiment is in direct contrast to the consensus of health officials that face-masks help combat the pandemic. This warranted the need to study the attitude of individuals towards face-masks expressed via various social networks.\\

Several recent studies have focused on the rise of anti-mask sentiments across social media, with particular focus on the platform Twitter. \cite{twitter} investigated the ``types, themes, temporal trends, and exchange patterns of hashtags about mask wearing" on Twitter. \cite{twitter2} analyzed tweets regarding the use of masks during the pandemic. \cite{twitter3} examined the relationship between the number of anti-mask tweets and the number of new daily cases of COVID-19 via Pearson correlation analysis between the two-time series. \cite{twittertoxicity} focused on the ``toxicity of discourse on Twitter" around wearing face masks and how it aids the spread of misinformation regarding the pandemic. They have identified hashtags that expressed pro-mask as well as anti-mask sentiments. \cite{russian} conducted a cross-sectional study to analyze the attitude towards regulations regarding face-masks on the Russian social network VKontakte.\\

To the best of the authors' knowledge, no study on mask-related sentiments has been carried out so far that focuses on the platform Reddit.

\subsection{The current work:}

In the preliminary stages of the study, the broad topics of interest were identified. Subsequently, the following concrete Research Questions (RQ) were formulated: 
\begin{enumerate}
	\item RQ1: What percentage of the population participating in a particular subreddit perceives face masks as a useful measure in preventing the spread of COVID-19, and what percentage of the population is committed to utilizing the same? 
	\item RQ2: What are the habits pertaining to the use of masks? This encompasses questions such as: 
	\begin{enumerate}
		\item Do people cover their mouths or noses or both?
		\item How many layers are there in a person's face-mask of choice? 
		\item How often do people wash their masks? 
	\end{enumerate} 
	\item RQ3: Do people use additional protective equipments such as face shields, gloves, or shower caps?
	\item RQ4: Are individuals with co-morbidities or those residing with someone who has co-morbidities in support of wearing masks? Do people who contracted COVID-19 wear masks?
	\item RQ5: What are the nuisances and inconveniences faced by people when wearing a mask?
	\item RQ6: What external opinions, if any, influence a person's decision to wear masks?
\end{enumerate}

An extensive review of literature was carried out by a group of researchers well-versed in survey sampling. Subsequently, an international online survey was designed, with English selected as its primary and only language. It was uploaded to the electronic survey platform provided by Google, and shared to the community /r/SampleSize in the popular social network Reddit.\\

To the best of the authors' knowledge, no attempts have been made to conduct such a comprehensive and detailed survey regarding an individual's beliefs, habits and complaints related to face-masks. The authors could not find any such survey that focused on Reddit. Therein lies the novelty of the manuscript. \\

The methods used for the survey have been detailed in Section (\ref{methods}). Section (\ref{demo}) presents the demographic information of the survey participants. The results of the survey have been presented and discussed in Section (\ref{results}), and the conclusions in Section (\ref{conclusions}). Section (\ref{crossval}) discusses the cross-validation of the results. Section (\ref{future}) highlights the future scope of study, while Section (\ref{strengths_limitations}) outlines the strengths and limitations of the study.

\section{Methods:}
\label{methods}

An international online survey was designed and uploaded to the survey platform provided by Google and subsequently shared to /r/SampleSize in Reddit. Details follow on the choice of the subreddit as the platform for the study, as well as the survey methods employed for the study.  

\subsection{The choice of Reddit as the social media platform for the study:}

As noted earlier in the Introduction section, social networks have a crucial role to play in advocacy and information exchange, specially during times of crisis. The alarming increase of anti-mask sentiments across the various platforms warrants a closer look into the mindsets of individuals partaking in such networks. This is the primary motivation behind the study.\\

The social network Reddit is a content-aggregator platform compromised of various communities called subreddits that contain user-generated content and discussions on a particular topic of interest. It  is currently one of the major social networks in use, alongside Facebook, Instagram, Twitter, etc, with over 430 million users as of April 2021. (Source: https://www.statista.com/ statistics/272014/global-social-networks-ranked-by-number-of-users/).\\

 Several studies related to COVID-19 with a focus on Reddit communities have been carried out recently, such as  \cite{reddit1}, \cite{reddit4}, \cite{reddit2}, \cite{reddit3}. \cite{reddit3} have highlighted the strength of Reddit as a source for understanding public reactions to the pandemic, by studying two communities in particular, namely, /r/Coronavirus, and /r/China\_flu. With these facts in mind, it is reasonable to choose Reddit as a platform for the study.\\

 To the best of the authors' knowledge, no study has been conducted to understand the perception and attitude of users of a particular subreddit towards the use of face-masks during the pandemic. 

\subsection{The choice of subreddit /r/SampleSize:}

For the purpose of the current study, the community /r/SampleSize was utilized. This subreddit is a community for Reddit users to post links to their surveys, as well as participate in surveys posted by other users. While any Reddit user can view such surveys, the best estimate of potential participant exposure is given by the number of users who are subscribed to the subreddit. At the time of the beginning of the survey, the subreddit had 158,000 subscribers, while at the time of closing, this grew to 161,000 subscribers.\\

The choice of this particular subreddit was motivated by the need to find a subreddit that contains a diverse group of users. Subreddits that deal specifically with COVID-19 were deliberately avoided to avoid bias and the ``echo chamber" effect previously mentioned, since some subreddits such as /r/CovIdiots are significantly pro-mask, while others like /r/NoNewNormal have a notable anti-mask stance. /r/SampleSize is a subreddit where COVID-19 is not the primary focus of discussions. No blatant opinions are expressed on the subreddit to indicate whether an individual user is pro-mask and anti-mask. Hence it is interesting to study the latent beliefs of the users.\\

Further, the subreddit provided a diverse group of respondents, which is evident from the Demographic Information of the study, presented in Section \ref{demo}. \cite{samplesize} have noted that quality and reliability of data from /r/SampleSize were comparable to lab samples across most analyses. They concluded that the subreddit contains a diverse and viable pool of survey participants. Hence, the choice of the subreddit /r/SampleSize for conducting the survey is justified.

\subsection{The use of convenience sampling:}

The users of Reddit are anonymous in nature and the demographic characteristics of an individual Reddit user are completely unknown.The absence of a sampling frame consisting of relevant information of all subscribers is an impediment to determining a criteria for inclusion or exclusion of a unit in the sample. Hence, it is infeasible to draw a sample via any probability sampling scheme. Therefore, use of convenience sampling is justified for this particular study.\\

Additionally, to overcome the shortcomings of convenience sampling, a second, smaller survey was conducted after a few months to collect data and cross-validate the original results. 

\subsection{Particulars of the survey:}

A total of 30 questions pertaining to the perception and attitude of the users towards masks were included in the questionnaire and presented in simple, unambiguous language. These questions covered a broad spectrum of topics, including demographic information, as well as the topics related to face-masks identified by the Research Questions 1-6.\\

Answering some fundamental questions was mandatory, whereas participants had the option to not disclose certain data that may be deemed sensitive, such as demographic information, health-related information, etc. A few questions were conditional on answering another question affirmatively. Most questions had binary (Yes/No) options or Multiple Choice answers, so that data collected was binary or categorical in nature. However, participants were given the liberty of adding their own answers or comments for some questions. \\

The survey was made available online for public participation on 24 February 2021. A link to the survey was distributed periodically on the subreddit /r/SampleSize. The survey was open to the global population, with the only restriction being the participant's ability to answer questions in English. A total of $1425$ responses was collected after a month, and the survey was closed on 23 March 2021. The present study focuses on the analysis of those responses.

\subsection{Data privacy and participation:}

In compliance with Google's privacy policy (https: $\backslash \backslash$policies.google.com$\backslash$privacy?hl=en), all responses were completely anonymous. No identifying information or contact information was solicited from the respondents. Further, participants could exit the survey at any moment by simply closing their browser window. Responses were recorded and stored only when the ``Submit" button was clicked. Participation was completely voluntary, and no monetary or other incentives were provided to the participants for partaking in the survey.

\section{Demographic information}
\label{demo}

The survey was successful in attracting a diverse pool of participants. The key demographic characteristics of the respondents have been tabulated in Table (\ref{demoinfo}) and discussed briefly in this section.

\begin{table}
	\caption{Demographic information}
	\label{demoinfo}
\begin{center}
\begin{tabular}{llrr}
	\hline 
	Variable & Range or Values & No. of responses & \% of responses \\
		\hline 
\multirow{7}{*}{Age} & $<18$ & 154 &	10.81\\
& 18-24 &	630 &	44.21\\
& 25-34	& 474	& 33.26\\
& 35-44 &	123	& 8.63\\
& 45-54	& 23	& 1.62\\
& $>$55	& 16	& 1.12\\
& Undisclosed &	5	& 0.35\\
	\hline 
\multirow{4}{*}{Gender identity} & Female &	900 &	63.16\\
	& Male &	465 &	32.63\\
	& Non-binary &	37 & 2.60\\
	& Undisclosed	& 23 & 1.61\\
\hline 
\multirow{8}{*}{Continent} & Africa &	7 &	0.49\\
& Asia &	61 &	4.28\\
& Asia/Europe &	8	& 0.56\\
& Europe	& 461	& 32.35\\
& North America	& 830	& 58.25\\
& Oceania	& 34	& 2.39\\
& South America	& 16	& 1.12\\
& Misc	& 8 &	0.56\\
\hline
\multirow{3}{*}{Nature of residential area} 
& Rural & 	166	& 11.65\\
& Suburban &	632	& 44.35\\
& Urban	& 627	& 44.00\\
\hline 
\multirow{6}{*}{Relationship status} & Divorced &	11 & 0.77\\
& Engaged	& 47	& 3.30\\
& In a relationship &  	422	& 29.61\\
& Married &	212	& 14.88\\
& Single &	722	& 50.67\\
& Undisclosed	& 11	& 0.77\\
\hline
\multirow{5}{*}{Living situation} & Alone  	& 173 &	12.14\\
& With family or partner	& 1032	& 72.42\\
& With one or more roommates  	& 218	& 15.30\\
& Special	& 1 &	0.07\\
& Undisclosed	& 1 &	0.07\\
\hline 
\end{tabular}
\end{center}
\end{table}

\begin{figure}[!ht]
	\subfloat[\label{A}]{\includegraphics[width=0.45\textwidth]{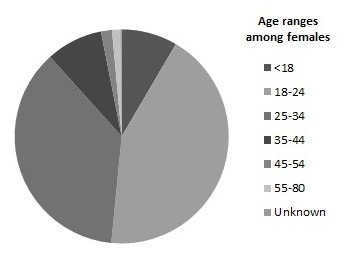}}
	\hfill
	\subfloat[\label{B}]{\includegraphics[width=0.45\textwidth]{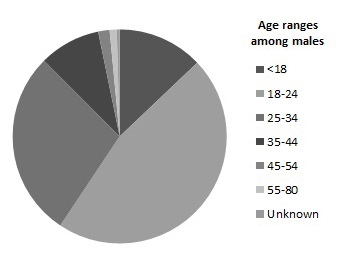}}
	\hfill
	\subfloat[\label{C}]{\includegraphics[width=0.45\textwidth]{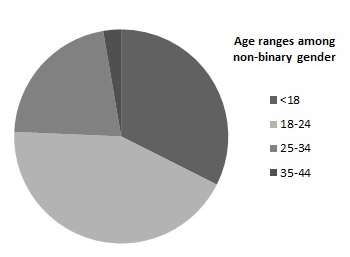}}
	\hfill
	\subfloat[\label{D}]{\includegraphics[width=0.45\textwidth]{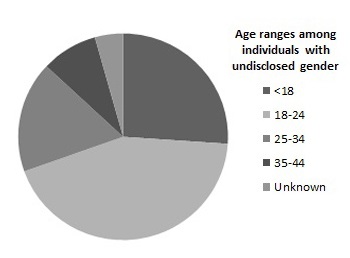}}
	\caption{Gender-wise age ranges of individuals}
	\label{genderage}
\end{figure}

\subsection{Age:} 
Majority of the participants (1381 out of 1425) were young or middle-aged (aged less than 45). Only 44 respondents aged 45 or older could be reached. This is in par with the demographics of people who actively use social networks, as seen by a Pew Research Centre study in 2018 (\cite{pew}). \\

The ages of the participants according to their gender identities have been represented by pie-charts in Figure (\ref{genderage}).

\subsection{Gender identity:}
Majority of the respondents (63.16\%) identified as female, while 32.63\% identified as male. The non-binary category included 13 individuals identifying as non-binary, as well as its synonyms, such as agender(7 respondents), genderqueer(2 respondents), enby(1 respondent), female/non-binary(1 respondent) and transmasculine(1 respondent). A person identifying as ``Male, soon to transition to female" was also placed in this category. Only 1.61\% or 23 respondents did not disclose their gender identity.

\subsection{Geographical location:}

Responses were received from all over the globe, spanning 67 countries, namely, Antigua, Argentina, Australia, Austria, Bahrain, Belgium, Benin, Bolivia,
Bosnia and Herzegovina, Brazil, Bulgaria, Canada, China, Colombia, Cyprus, Czech Republic, Czechia, Denmark, Ecuador, Egypt, Finland, France, Georgia, Germany, Greece, Hong Kong, Hungary, Iceland, India, Ireland, Israel, Italy, KSA, Kuwait, Latvia, Lithuania, Malaysia, Mexico, Netherlands, New Zealand, North America, Norway, Pakistan, Philippines, Poland, Portugal, Puerto Rico, Romania, Russia, Serbia, Singapore, Slovakia, South Africa, South Korea, Spain, Sweden, Switzerland, Taiwan, Thailand, Turkey, UAE, Ukraine, Uruguay, USA, Venezuela and Yemen.\\

Majority of the respondents (58.25\%) were from North America, with  responses from the United States of America and Canada comprising respectively  50.46\% and 7.51\% of the total responses. Respondents from Europe constituted the second most prominent group(32.35\% responses), with the United Kingdom, Germany and the
Netherlands contributing 12.35\%, 4.28\% and 3.65\% of the responses respectively.\\

Some respondents were from countries that spanned across Asia and Europe, namely, Georgia, Russia, and Turkey, and were placed in the Asia/ Europe category. The ``Misc" category included respondents who listed two countries as their location (4 respondents), as well as those who chose not to disclose their location (4 respondents). 

\subsection{Classification of residential area (Urban, suburban, rural):}

The survey participants were mostly from a suburban area (44.35\%) or an urban area (44.00\%), with only 11.65\% of responses from rural areas.

\subsection{Current living situation:}

A total of 173 people reported living alone.\\

The category ``With family or partner" included 1032 respondents who lived with their parents or other family members, their partners (boyfriend/girlfriend, fianc\'{e}, wife/husband, domestic partner), or with their partners' family. People who typically live alone or in a dormitory/ student housing but were living with their family at the time of the survey were also placed in this category. \\

The category ``With one or more roommates" consisted of 218 respondents who reported living with a housemate, one or more roommates, or at a student residence/ dormitory.\\

One participant who reported being in none of the above categories was placed in the ``Special" category. The living situation of one respondent remained undisclosed.

\subsection{Relationship status:}

The survey was successful in reaching individuals who were in a relationship (422 responses), engaged (47 responses), married (212 responses), single (722 responses) and divorced (11 responses).

\section{Results:}
\label{results}

All statistical analyses for the purpose of the study were performed using the open source statistical software \texttt{R} (\cite{r}). Statistical significance was accepted as $p<0.05$. Binomial as well as multinomial confidence intervals were constructed for responses related to various variables of interest. The functions \texttt{BinomCI} and \texttt{MultinomCI} from \texttt{DescTools} \cite{desctools} package in \texttt{R} \cite{r} were utilized for the binomial and the multinomial cases respectively. Pearson's Chi-squared test was utilized to test whether a relationship exists between two variables of interest.\\

A detailed discussion of the results of the data analysis  follows.

\subsection{Mask use adherence and belief in the effectiveness of masks:}

The first focus of discussion involves RQ1: What percentage of the population participating in a particular subreddit perceives face masks as a useful measure in preventing the spread of COVID-19, and what percentage of the population is committed to utilizing the same?\\

Among the 1425 respondents surveyed, 1378 respondents or 96.70\% reported utilizing personal protective gear in the form of masks, while 47 respondents or	3.30\% reported not using them. Applying the \texttt{BinomCI} function from \texttt{DescTools} (\cite{desctools}) package in \texttt{R} (\cite{r}), confidence intervals for proportion of individuals who self-reported mask use adherence were constructed using various well-known methods (\cite{agresticoull, blaker,  brown, newcombe, pratt, wilcox, witting}). The results have been presented in Table (\ref{maskCI}). From the table, one can claim with 95\% confidence that the proportion of individuals in the target population who wear masks lies between 0.96 and 0.98.\\

Some respondents self-reported declining the use of masks; 19 of them (40.43\%) were from the USA, the country that has the highest number of confirmed COVID-19 cases in the world, while 5 respondents (10.64\%) were from the UK, another country that has been severely affected by COVID-19. A few respondents were from Australia (4 responses)  and New Zealand (5 responses), where the daily new cases were less than 15 during the time period when the survey was conducted. \\

\begin{table}[htb]
	\caption{Confidence intervals for proportion of individuals who self-reported mask use adherence}
	\label{maskCI}
\begin{center}
\begin{tabular}{lccc}
	\hline 
Method	& Estimate &	Lower CI &	Upper CI\\
\hline 
Wilson &	0.9670175 	& 0.9564171 & 0.9751068\\
Wald &	0.9670175 &	0.9577450 &	0.9762901\\
Waldcc &	0.9670175 &	0.9573941 &	0.9766410\\
Wilsoncc &	0.9670175 &	0.9560203 &	0.9754096\\
Agresti-coull &	0.9657620 &	0.9563334 &	0.9751905\\
Jeffreys &	0.9670175 &	0.9567773 &	0.9753645\\
Modified wilson &	0.9670175 &	0.9564171 &	0.9751068\\
Modified jeffreys &	0.9670175 &	0.9567773 & 	0.9753645\\
Clopper-pearson	& 0.9670175 &	0.9563799 &	0.9756671\\
Arcsine	& 0.9667719 &	0.9568412 &	0.9754445\\
Logit	& 0.9670175 &	0.9563748 &	0.9751314\\
Witting	& 0.9670175 &	0.9583610 &	0.9740068\\
Pratt	& 0.9670175 &	0.9579045 &	0.9756691\\
Midp	& 0.9670175 &	0.9567328 &	0.9754159\\
Lik	& 0.9670175 &	0.9568958 	& 0.9754672\\
\hline 
\end{tabular}
\end{center}
\end{table}

\begin{table}
	\caption{Belief regarding the effectiveness of masks in checking the spread of a pandemic}
	\label{belief}
\begin{center}
\begin{tabular}{lrrrr}
	\hline 
	& \multicolumn{2}{c}{People who wear masks} & \multicolumn{2}{c}{People who do not wear masks}\\
	Response & No. of responses & \% of responses & No. of responses & \% of responses \\
	\hline 
	Yes	& 1070 &	77.65 & 8	& 17.02\\
	Yes, but only to & \multirow{2}{*}{282}	& \multirow{2}{*}{20.46} & \multirow{2}{*}{18} &	\multirow{2}{*}{38.30}\\	
	a certain extent & & & & \\
	No &  	26 &	1.89 & 21 &	44.68\\
	\hline 
\end{tabular}
\end{center}
\end{table}

Statistics regarding the trust of people in the effectiveness of masks to reduce the spread of the disease have been presented in Table (\ref{belief}). Pearson's Chi-squared test is used in \texttt{R} to test whether there is a significant relationship between a person's belief in the efficacy of masks and a person's decision to utilize masks. The values obtained are: $$\text{X-squared} = 281.31, df = 2, \text{p-value} < 2.2e-16.$$ Since the p-value is less than the significance level of $0.05$, we conclude that the two variables are in fact dependent. In other words, mask use adherence is directly related to the personal belief in the efficacy of masks. 

\subsection{Habits and beliefs pertaining to the use of masks:}

This section is concerned with exploring RQ2: What are the habits pertaining to the use of masks? The survey respondents were instructed to self-report whether they believed they were wearing masks correctly. Additionally, a variety of questions were included to assess whether there were common transgressions, such as leaving either the nose or the mouth uncovered, removal of masks during conversations, etc. Questions pertaining to the sanitation of reusable masks as well as the specifics of masks, such as the number of layers, were also posed.

\subsubsection{Self-reported beliefs regarding correctness of wearing masks:}

Of the 1378 respondents who self-reported complying with the recommendation of wearing masks, 1330 (96.52\%) believed that they wore masks correctly. 1328 (99.85\%) of them reported wearing masks that covered both their noses and their mouths, while 2 of them chose to cover only their noses with masks. 5 (0.38\%) respondents reported regularly removing their masks during conversations with someone, 92 (6.92\%) respondents reported doing so occasionally, while 1233 (92.70\%) claimed to never do so. \\

The remaining 48 (3.48\%) respondents believed that they did not wear masks correctly. Of them, 42 respondents reported that both their mouths and their noses were covered by masks, while 6 of them wore masks that covered only their mouths. 4(8.33\%) respondents admitted to the habit of regularly removing their masks when conversing with someone, 12 (25\%) respondents reported occasionally doing so, while 32 (66.67\%) reported never doing so.\\

Applying  \texttt{MultinomCI} function from \texttt{DescTools} (\cite{desctools}) package in \texttt{R} (\cite{r}), the 95\% confidence intervals for complete coverage of face by masks, occasional removal of masks, and regular removal of masks during conversations have been determined. The results have been presented in Table (\ref{maskremoveCI}). One can claim with 95\% confidence that a proportion of 0.927 of the surveyed individuals do not remove their masks during conversations.

\begin{table}
\caption{Multinomial Confidence Intervals for non-removal of face-masks, occasional removal of masks, and regular removal of masks during conversations}
\label{maskremoveCI}
\begin{center}
\begin{tabular}{ccccccc}
	\hline
	& \multicolumn{3}{c}{Among individuals who believed} & \multicolumn{3}{c}{Among individuals who believed}\\
	& \multicolumn{3}{c}{they wore masks correctly} & \multicolumn{3}{c}{they did not wear masks correctly}\\
		\hline
Response & Estimate & Lower CI & Upper CI & Estimate & Lower CI & Upper CI \\
	\hline
Regular removal & 0.003759398 & 0.00000000 & 0.01748400 & 0.08333333 & 0.0000000 & 0.2115613\\
Occasional removal & 0.069172932 & 0.05639098 & 0.08289754 & 0.25000000 & 0.1250000 & 0.3782280\\
Non-removal & 0.927067669 & 0.91428571 & 0.94079228 & 0.66666667 & 0.5416667 & 0.7948946\\
	\hline
\end{tabular}
\end{center}
\end{table}

\subsubsection{Mask cleaning and sanitation:}

Most respondents reported washing their masks daily (17.34\%), once in 2-3 days (16.47\%), or once a week (24.38\%).  442 respondents or 32.08\% reported wearing disposable masks. The detailed responses relating to the frequency with which the respondents reported washing their masks has been presented in Table (\ref{frequencywash}). The Table also contains the 95\% confidence intervals for the proportion of individuals in the population for the same, attained by employing \texttt{MultinomCI} function from \texttt{DescTools} (\cite{desctools}) package in \texttt{R} (\cite{r}).\\

Some notable responses received in the ``Other" category includes using multiple masks and bulk-washing them at the end of a week or two weeks, using masks that required cleaning by ultra-violet (UV) filtration instead of washing, simultaneous use of disposable and reusable masks for the purpose of double-masking, washing masks after every use, washing masks every 4-6 hours, washing masks every two weeks, and washing masks solely if they became soiled or putrid. Some reported reusing disposable masks. Some also noted that their infrequent washing of masks was a reflection of their sporadic need to venture out of their homes. 

\begin{table}
\caption{Frequency of washing of masks}
\label{frequencywash}
\begin{center}
\begin{tabular}{lrrcc}
	\hline 
Frequency of washing & No. of responses & \% of responses & Lower CI & Upper CI\\
\hline 
Daily  & 239 &	17.34 & 0.1465893 & 0.20162869\\
I wear disposable masks  &	442	& 32.08 & 0.2939042 & 0.34894364\\
Once in 2-3 days  &	227 &	16.47 & 0.1378810 & 0.19292042\\
Once a week  &	336 &	24.38 & 0.2169811 & 0.27202056\\
Once a month  & 	10	& 0.73 & 0.0000000 & 0.03544582\\
Rarely   &	15 &	1.09 & 0.0000000 & 0.03907426\\
Never   &	15	& 1.09 & 0.0000000 & 0.03907426\\
Other   &	94	& 6.82 & 0.0413643 & 0.09640373\\
\hline
\end{tabular}
\end{center}
\end{table}

\subsubsection{Number of layers in masks:}

The respondents were asked to report the number of layers in their masks. The most commonly used masks contained two layers (36.43\%), followed by three layers (21.84\%). A significant percentage of the respondents (21.92\%) were unaware of the number of layers in their masks.\\

The responses, as well as the 95\% confidence intervals for the proportion of individuals in the population with such responses, obtained using \texttt{MultinomCI} function from \texttt{DescTools} (\cite{desctools}) package in \texttt{R} (\cite{r}), have been presented in Table (\ref{layers}). 

\begin{table}
	\caption{Layers in masks}
	\label{layers}
\begin{center}
\begin{tabular}{lrrcc}
	\hline 
No. of layers & No. of responses & \% of responses & Lower CI & Upper CI \\
\hline 
Four or more  &	39 & 2.83 & 0.0007256894 & 0.05682763\\
Three  & 301 & 21.84 & 0.1908563135 & 0.24695825\\
Two  & 502	& 36.43 & 0.3367198839 & 0.39282182\\
One  &	234	& 16.98 & 0.1422351234 & 0.19833706\\
I don't know  &	302	& 21.92 & 0.1915820029 & 0.24768394\\
\hline
\end{tabular}
\end{center}
\end{table}

\subsection{Use of additional personal protective equipments, such as face shields, gloves and shower caps:}

The next topic of interest is RQ3: Do people use face shields, gloves, or shower caps?\\

Most respondents reported not availing additional equipment, such as face shields (89.43\%), gloves (78.01\%) or shower caps (99.41\%). A summary of the responses related to the same has been compiled in Table (\ref{additional}). 

\begin{table}
	\caption{Use of additional protective equipments, such as face shields, gloves and shower caps}
	\label{additional}
\begin{center}
\begin{tabular}{lrrrrrr}
	\hline
	& \multicolumn{2}{c}{Face shields} & \multicolumn{2}{c}{Gloves} & \multicolumn{2}{c}{Shower caps} \\
&	No. of 	& \% of  &	
No. of & \% of  & No. of & \% of  \\
Response & responses & responses & responses & responses & responses & responses\\
\hline 
No &	1233	& 89.48	& 1075	& 78.01	& 1370	& 99.41\\
Occasionally &	95	& 6.89	& 261	& 18.94	& 4	& 0.29\\
Yes	& 47 & 3.41	& 41 & 2.98	& 2	& 0.15\\
Unknown & 3	& 0.22	& 1	& 0.07	& 2	& 0.15\\
\hline 
\end{tabular}
\end{center}
\end{table}

\subsection{Mask use adherence and comorbidities:}

Individuals having co-morbidities such as compromised immune system, heart disease, lung disease, obesity, diabetes, etc. are at a greater risk of grave complications from COVID-19 (\cite{comorbidity}). Hence it is particularly crucial that they follow personal safety protocols and minimize contact with individuals who flout such protocols. The focus of this section is RQ4: Do individuals with co-morbidities or those residing with someone having co-morbidities wear masks? Do people who contracted COVID-19 wear masks?

\subsubsection{Prevalence of mask use among individuals with co-morbidities:}

\begin{figure}[ht]
	\centering
	\subfloat[\label{A1}]{\includegraphics[width=0.3\textwidth]{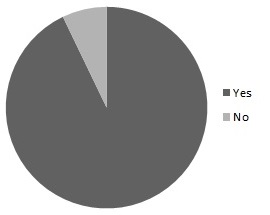}}
	\subfloat[\label{B1}]{\includegraphics[width=0.3\textwidth]{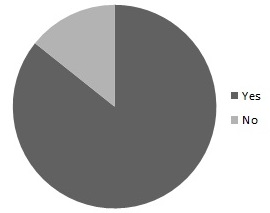}}
	\subfloat[\label{C1}]{\includegraphics[width=0.3\textwidth]{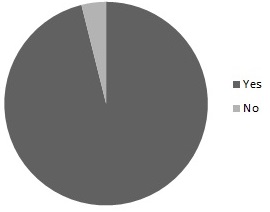}}\\
	\subfloat[\label{D1}]{\includegraphics[width=0.3\textwidth]{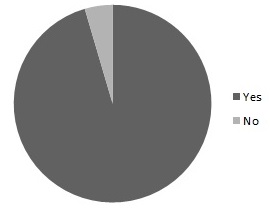}}
	\subfloat[\label{E1}]{\includegraphics[width=0.3\textwidth]{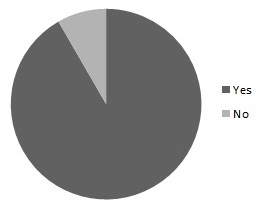}}
	\caption{Response (Yes/No) to the question ``Do you wear masks?" among a) immuno-compromised individuals b) individuals with heart disease c) individuals with long disease d) individuals with obesity e) individuals with diabetes}
	\label{comorbfig}
\end{figure}

Table (\ref{comordities}) presents some statistics on the use of masks among those with the above-mentioned co-morbidities. It is seen that 92.86\% of respondents who were immuno-compromised, 71\% of those with heart diseases, 96\% of those with lung diseases, 95.45\% of those with obesity, and 91.67\% of those with diabetes self-reported wearing masks. This has been represented by pie-charts in Figure (\ref{comorbfig}). \\

\begin{table}
	\caption{Prevalence of wearing masks among those with co-morbidities}
	\label{comordities}
\begin{center}
\begin{tabular}{lrrrrrr}
	\hline 
	& \multicolumn{3}{c}{Responses among mask-wearers} &	\multicolumn{3}{c}{Responses among non-mask wearers}\\
	\hline 
Categories	& Yes &	No & Undisclosed & Yes & No	& Undisclosed\\
\hline 
Immuno-compromised & 91 & 1260 & 27 & 7 & 37&3\\
Heart disease &	18 & 1353 & 7 &	3 & 44 & 0\\
Lung disease & 74 &	1297 & 7 & 3 & 44 & 0\\
Obsesity & 168 & 1198 &	12 & 8	& 38 & 1\\
Diabetes &	22	& 1348 & 8	& 2	& 44 & 1\\
\hline
\end{tabular}
\end{center}
\end{table}

\subsubsection{Prevalence of mask use among individuals residing with people having co-morbidities:}

Some statistics about usage of masks among people who share a living space with one or more individuals having the above-mentioned co-morbidities has been presented in Table (\ref{cohabitcomor}). Of the 391 individuals who reported residing with someone having one or more of the co-morbidities, 379 individuals or 96.93\% reported using masks. 

\begin{table}
	\caption{Prevalence of mask use among those residing with others having co-morbidities}
	\label{cohabitcomor}
\begin{center}
\begin{tabular}{crrrr}
	\hline 
&	Lives with someone  &	Does not live with someone  &	Undisclosed	& Lives alone \\
& with co-morbidities & with co-morbidities & & \\
	\hline 
Wear Masks &	379	& 884 & 14	& 101\\
Do not wear masks &	12	& 27	& 1	& 7\\
	\hline 
\end{tabular}
\end{center}
\end{table}

\subsubsection{Prevalence of mask use among those with COVID-19:}

Table (\ref{covidmask}) presents information on whether a respondent contracted COVID-19 and whether they use masks. It is to be noted that no information was collected on whether individuals contracted COVID-19 despite wearing masks, or whether they started wearing masks after contracting COVID-19. This may be the subject of future investigations, and was included as a question in the cross-validation survey.

\begin{table}
	\caption{Prevalence of masks among those with Covid-19}
	\label{covidmask}
	\begin{center}
		\begin{tabular}{crr}
			\hline 
			& No. of respondents & No. of respondents\\
			& Who wear masks &	Who do not wear masks\\
			\hline 
			Did not contract Covid-19 &	1259	& 40\\
			Contracted Covid-19	& 110 &	6\\
			\hline 
		\end{tabular}
	\end{center}
\end{table}

\subsection{Nuisances and inconveniences faced by people in relation to masks:}

The next focus of discussion is RQ5: What are the nuisances and inconveniences faced by people when wearing a mask? Survey respondents were questioned on the same. The responses have been tabulated in Table (\ref{problems}). The 95\% confidence intervals for the proportion of individuals in the population facing such problems was obtained using \texttt{BinomCI} function from \texttt{DescTools} (\cite{desctools}) package in \texttt{R} (\cite{r}). This data has also been presented in the table.\\

The most common complaints were difficulty in communicating with people when wearing masks (58.39\%), pain in earlobes from prolonged use of masks with strings (37.89\%), difficulty breathing (28.84\%) and excessive perspiration (24.28\%).\\

\begin{figure}[ht]
	\centering
	\subfloat[\label{gl}]{\includegraphics[width=0.45\textwidth]{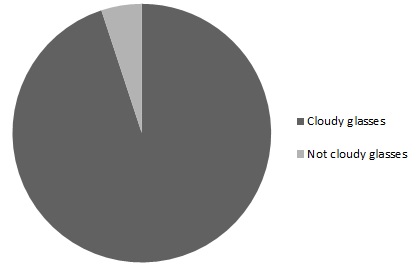}}
	\subfloat[\label{ch}]{\includegraphics[width=0.45\textwidth]{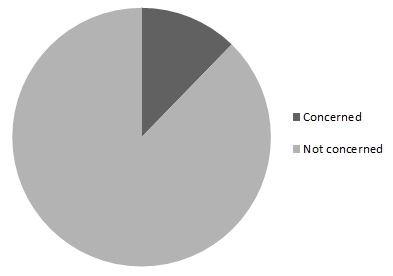}}
	\caption{Responses of individuals regarding a) glasses becoming cloudy due to masks b) concerns for safety of small children when strangers wear masks}
	\label{glasseschildren}
\end{figure}

Additionally, of the 870 respondents who reported wearing spectacles, a majority of 826 individuals or 94.94\% had grievances about masks clouding up the glasses. Of the 220 people with young children, only 27 (12.27\%) were apprehensive about the safety of their children around strangers wearing masks, while 193 (87.73\%) had no such concerns. This has been represented by pie-charts in Figure (\ref{glasseschildren}).

\begin{table}
	\caption{Nuisances and inconveniences faced by people in relation to masks}
	\label{problems}
\begin{center}
\begin{tabular}{lrrcc}
	\hline 
& No. of & \% of && \\
Problems & responses  & responses & Lower CI & Upper CI\\
\hline 
Difficulty breathing 	& 411 &	28.84 & 0.2654931 & 0.3124867\\
Earache from the mask strings &	540	& 37.89 & 0.3541166 & 0.4044291\\
Excessive perspiration &	346 &	24.28 & 0.2212551 & 0.2657418\\
Anxiety and/or feeling of suffocation & 164	& 11.51 & 0.0995432 & 0.1327019\\
Difficult to hear people and vice versa &	832	& 58.39 & 0.5580750 & 0.6091934\\
Impediment of fashion statement & 76	& 5.33 & 0.0428217 & 0.0662467\\
Concerns about safety of public places & \multirow{2}{*}{45} & \multirow{2}{*}{3.16} & \multirow{2}{*}{0.0236838} &  \multirow{2}{*}{0.0419928}\\
when strangers wear masks &  & & & \\
None of the above &	299	& 20.98 & 0.1894775 & 0.2317319\\
\hline 
\end{tabular}
\end{center}
\end{table}

\subsection{Influence of other individuals on the decision to wear masks:}

The final item of discussion involves RQ6: What external opinions, if any, influence a person's decision to wear masks?\\

The survey respondents were questioned on whether people in their personal circle, such as family members, friends, doctors, etc. had urged them to use masks. The responses and the corresponding 95\% confidence intervals for the proportion of individuals in the population with such responses, obtained using \texttt{BinomCI} function from \texttt{DescTools} (\cite{desctools}) package in \texttt{R} (\cite{r}), have been presented in Table (\ref{personal}).\\

The respondents were also asked to identify the public figures that they had seen advocating for the use of masks. Table (\ref{public})
 contains the responses and the corresponding 95\% confidence intervals obtained using \texttt{BinomCI} function from \texttt{DescTools} (\cite{desctools}) package in \texttt{R} (\cite{r}).\\
 
 Table (\ref{maskers_influence}) presents statistics on whether a respondent reported that they would be more comfortable wearing masks if their political or religious leaders promoted its use. Pearson's Chi-squared test is used in \texttt{R} to test whether there is a relationship between religious or political figures urging a person to wear masks and a person's decision to wear masks. The values obtained when analyzing political influence were: $$\text{X-squared} = 25.547, df = 2, \text{p-value} = 2.835e-06,$$ whereas the values obtained during analysis of religious influence were: $$\text{X-squared} = 25.752, df = 2, \text{p-value} = 2.558e-06.$$

Since the p-value is less than the significance level of $0.05$ in both the cases, we conclude that people would be more likely to don masks if their political or religious leaders urged them to do so.

\begin{table}
	\caption{Statistics on which individuals in a person's circle had urged them to wear masks}
	\label{personal}
\begin{center}
\begin{tabular}{lrrcc}
\hline 
\multirow{2}{*}{Category} & No. of & \% of & \multirow{2}{*}{Lower CI }& \multirow{2}{*}{Upper CI}\\
 & responses & responses & &  \\
\hline 
A family member &	782	& 54.88 & 0.5228387 &  0.5744429\\
A close friend	& 549 &	38.53 & 0.3603361 & 0.4108071\\
Significant other & 349	& 24.49 & 0.2232898 & 0.2679063\\
Doctor	& 290 &	20.35 & 0.1834151 & 0.2251967\\
Colleague, teacher, friend, etc. &	518	& 36.35 & 0.3389323 & 0.3888192\\
A stranger	& 166 &	11.65 & 0.1008559 & 0.1341887\\
None of the above &	467	& 32.77 & 0.3038402 & 0.3525247\\
\hline 
\end{tabular}
\end{center}
\end{table}

\begin{table}
	\caption{Statistics on the public figures that respondents have noticed urging people to wear masks}
	\label{public}
\begin{center}
\begin{tabular}{lrrcc}
	\hline 
	\multirow{2}{*}{Category} & No. of & \% of & \multirow{2}{*}{Lower CI} & \multirow{2}{*}{Upper CI}\\
	& responses & responses & & \\
	\hline
	Local government &	1198 &	84.07 & 0.8207886 & 0.8587829\\
	President or Prime Minister &	1059 &	74.32 & 0.7198415 & 0.7651668\\
	A political leader from the party they support & 936	& 65.68 & 0.6317999 & 0.681041\\
	A political leader from the opposition party	& 508 &	35.65 & 0.3320394 & 0.3817147\\
	A religious leader &	250	& 17.54 & 0.1565708 & 0.1960515\\
	A celebrity they like &	862	& 60.49 & 0.5792803 & 0.6299801\\
	A celebrity they do not like & 575	& 40.35 & 0.3783288 & 0.4292076\\
	None of the above &	71 & 4.98 & 0.0396883 & 0.06238143\\
	\hline
\end{tabular}
\end{center}
\end{table}

\begin{table}
	\caption{Statistics on whether individuals would be more comfortable wearing masks if their political or religious leaders urged them}
	\label{maskers_influence}
\begin{center}
\begin{tabular}{lcccccc}
	\hline 
	& \multicolumn{3}{c}{No. of responses among} &
	 \multicolumn{3}{c}{No. of responses among}\\
	 & \multicolumn{3}{c}{individuals who wear masks} & \multicolumn{3}{c}{individuals who do not wear masks}\\
&	Yes	& No &	Not political/ religious &	Yes	& No &	Not political/ religious \\
\hline 
Political leaders	& 443	& 483	& 499 & 3	& 32 & 12\\
Religious leaders&	148 & 289 &	988 &	3 &	24 & 20\\
\hline 
\end{tabular}
\end{center}
\end{table}

\section{Cross-validation of results:}
\label{crossval}

Since the sample was collected via convenience sampling, a second, smaller sample was drawn to cross-validate the results. The questionnaire from the previous survey was retained, with two modifications: 

\begin{enumerate}
	\item At the time the second survey was conducted, vaccinations were in full swing in many countries and mask-mandates were reflected. Hence, respondents were given a third option to the question  ``Do you wear masks?"  besides ``Yes" and ``No" to indicate that they complied with mask-regulations before vaccination. Further, they were instructed to respond to the sections regarding masks based on their experience during the time they used the same. 
	\item An additional question was included to ask whether individuals who contracted COVID-19 did so despite wearing masks, whether they did not wear masks before the disease but wear them after, or whether they did not wear masks when contracting the disease and do not do so after. 
\end{enumerate}

 The survey was circulated during the third week of July, 2021. A total of 300 responses were collected. The summary of the results have been tabulated and presented as Supplementary Material.\\ 
 
 Although the exact numbers and percentages varied slightly between the two surveys, the results remained consistent and the conclusions were the same. This indicates that despite the use of convenience sampling, the sample collected provides a good estimate of the characteristics of the target population. 

\section{Conclusions:}
\label{conclusions}

Despite vocal anti-mask protests both on the streets and on social networks, an overwhelming majority of the respondents reported believing that masks are efficient in reducing the spread of the COVID-19 disease. The majority also reported adhering to the use of masks. This gives us an instance of at least one subreddit where the majority do not have an anti-mask stance. The study demonstrates that despite the digital polarization of public opinion on using face-masks to battle COVID-19, the prevalence of opinions on the matter is highly unbalanced in a forum where COVID-19 is not the primary focus of conversation. The pro-mask stance of /r/SampleSize suggests one should be cautious about erroneously assuming that opinions from both sizes of the spectrum are ubiquitous.\\

Additionally, the results of the study have potential for applications in improving the battle against COVID-19 and in battling future pandemics. Some additional conclusions that may be drawn with regards to the view of members of the subreddit /r/SampleSize about masks are as follows:

\begin{enumerate}
\item Most of the individuals with co-morbidities or those living with someone with co-morbidities reported complying with the guidelines of wearing masks.
\item The responses indicate that the governments were able to relay the message of wearing masks at both the local and the federal levels. Political leaders were also successful in conveying such messages. A relatively small percentage of participants reported seeing such messages from religious leaders. 
\item The responses also indicate that individuals would be more comfortable wearing masks if their political or religious leaders urged them to do so. 
\end{enumerate}

The results of the study may help the governments better understand the perspectives of individuals regarding masks and the problems they face in its use. Based on the responses obtained during the course of the study, some areas that may benefit from suitable changes were identified. The following recommendations may be made to the concerned authorities:

\begin{enumerate}
	\item Increasing public awareness: 
\begin{enumerate}
\item There is a need for increased public awareness regarding the disposal of disposable masks and sanitation of reusable ones. 
\item Although the majority of the individuals reported wearing masks that covered both their mouths and their noses, further awareness is needed, as some tend to leave either uncovered.
\item Use of additional protective equipments such as face shields, gloves and shower caps should be encouraged in public places like restaurants, malls, etc. after reopening.
\end{enumerate}
\item Improving the experience of wearing masks for the general public:
\begin{enumerate} 
\item More resources should be devoted by manufacturers to develop masks out of fabrics that minimize perspiration and make them more breathable, while simultaneously providing adequate protection from viruses. 
\item Masks that utilize strings behind earlobes were reported to cause pain, hence innovations are needed to develop alternative mechanisms. 
\item Innovative technologies are also needed to prevent clouding of spectacles when one wears masks. 
\end{enumerate}
\end{enumerate}

\section{Future scope of study:}
\label{future}

There is scope for further research on digital polarization on salient issues in social media. As communication processes evolve, research on the same should evolve accordingly. Considering the fact that Reddit is currently one of the major social networks, there is ample scope for utilization of Reddit as a platform for research on digital polarization of opinions.\\

The study can be expanded in the future by targeting other subreddits, and including subreddits with a primary focus on COVID-19, to compare and contrast the opinions of users among various subreddits. The results of the study imply that users with vocal anti-mask sentiments do not actively participate in unrelated subreddits; this may be further explored.\\ 

Other areas of interest include targeting people from specific countries and analyzing country-specific data to understand the localised opinions, follow-up surveys that take place in the intervals of a few months to cross-validate the data, etc.

\section{Strengths and limitations of the study:}
\label{strengths_limitations}

The strength of the survey includes its ability to reach a diverse range of  participants in several continents at minimal cost due to the survey being entirely online. Survey participation was completely anonymous and voluntary, hence it is reasonable to believe that the responses received are honest. The survey targeted social media users, which was apt, considering the vocal protests against masks in various social network websites.\\

There were also limitations due to the electronic nature of the survey. There were no defined inclusion- exclusion criteria and no scope to select narrow target groups to form a representative sample. The validity of answers can not be verified. These are problems associated with online surveys and not exclusive to the current study.\\

Despite the limitations of the research, this study provides a snapshot of public opinion on the use of face-masks in the wake of Covid-19 among the users of a particular subreddit. The results identify at least one subreddit where pro-mask stance is dominant and highlight the scope for future research in the area. 

\section*{Disclosure statement:}

The authors have no conflicts of interest to disclose.

\section*{Data availability:}

The corresponding author may be contacted for the associated datasets. 

\section*{Funding:}

We are thankful to the Department of Science and Technology, Science \& Engineering Research Board (DST-SERB) for providing financial assistance under Grant EMR/2017/000882.

\end{document}